\documentclass[showpacs,preprintnumbers,amsmath,amssymb,twocolumn,superscriptaddress,prb]{revtex4}
\usepackage{graphicx}
\usepackage{amsfonts}
\usepackage{amsmath, amsthm, amssymb}
\usepackage{dsfont}
\usepackage{times}
\usepackage{subfigure}

\newcommand{\ve}[1]{\mathbf{#1}}
\newcommand{\vk}{\ve{k}} 

\newcommand{\e}[1]{\mathrm{e}^{#1}}

\newcommand{\kk}{\mathbf{k}}
\newcommand{\dd}{\mathbf{d}}

\newcommand{\bq}{\begin{equation}}
\newcommand{\eq}{\end{equation}}
\newcommand{\eps}{\varepsilon}
\newcommand{\state}[1]{|#1\rangle}

\newcommand{\eg}{\textit{e.g. }}
\newcommand{\etal}{\emph{et al.}}
\def\i{\mathrm{i}}

\begin{document}
\title[Nontrivial interplay between superconductivity and spin-orbit
coupling in non-centrosymmetric ferromagnets]{Nontrivial interplay between superconductivity and spin-orbit
coupling in non-centrosymmetric ferromagnets}
\author{Jacob Linder}
\affiliation{Department of Physics, Norwegian University of
Science and Technology, N-7491 Trondheim, Norway}
\author{Andriy H. Nevidomskyy}
\affiliation{Department of Physics and Astronomy,
Rutgers University, Piscataway, N. J., 08854-8019\\}
\author{Asle Sudb{\o}}
\affiliation{Department of Physics, Norwegian University of
Science and Technology, N-7491 Trondheim, Norway}

\date{Received \today}
\begin{abstract}
\noindent Motivated by the recent discoveries of ferromagnetic and non-centrosymmetric superconductors, we 
present a mean-field theory for a superconductor that \textit{both} lacks inversion symmetry and displays ferromagnetism, a scenario which is believed to be realized in UIr. We study the interplay between the order parameters to clarify how superconductivity is affected by the presence of 
ferromagnetism and spin-orbit coupling. One of our key findings is that the spin-orbit coupling seems to enhance both ferromagnetism and superconductivity in all spin channels. We discuss our results in the context of
the heavy fermion superconductor UIr and analyze possible symmetries of the order parameter by the  group theory method.
  \end{abstract}

\maketitle

In the past decade, a number of superconductors have been
discovered that are called `unconventional' as they fall outside the Bardeen-Cooper-Schrieffer (BCS) 
paradigm of electron-phonon mediated pairing with an isotropic gap. Of those, UPt$_3$ \cite{upt} and 
Sr$_2$RuO$_4$ \cite{sruo} were the first compounds to have been confirmed as triplet $p$-wave superconductors. 
More recently, a ferromagnetic (FM) superconductor was discovered in UGe$_2$ under 
pressure~\cite{saxena}, where the presence of an internal FM moment strongly suggests that only the 
equal-spin triplet pairing survives. In this latter example both the time-reversal and the gauge
symmetry due to SC order are spontaneously broken, which made UGe$_2$, as well as its cousins 
URhGe~\cite{URhGe} and UCoGe~\cite{UCoGe} an exciting avenue for theoretical and experimental 
research.
\par
For spin-triplet pairing, Anderson noticed~\cite{anderson} that
inversion symmetry is required to obtain a pair of
degenerate states $c^\dag_\kk \state{0}$ and  $c^\dag_{-\kk}
\state{0}$ capable of forming a Cooper pair. It was therefore 
surprising that superconductivity was discovered in the heavy
fermion compound CePt$_3$Si which lacks inversion symmetry~\cite{bauer}. It soon
became clear however that in the case of a non-centrosymmetric crystal, the
spin-orbit coupling (SOC) mixes different spin states, so that the
division into triplet and singlet symmetry of the SC order parameter
becomes meaningless. A bulk of theoretical work exists that has provided a
symmetry-based phenomenology to explain this in 
details~\cite{volovik87, shimahara, gorkov-rashba, frigeri04,samokhin_CePt3Si}.
The symmetry of the superconducting (SC) gap in this and other unconventional 
superconductors is presently a matter of intense investigation 
\cite{nelson,yuan,lebed,mazin}.
\par
An intriguing question  is what happens if time-reversal symmetry is broken in 
a crystal that lacks a centre of inversion. Can such a material become a superconductor? 
This question was answered affirmatively when superconductivity was discovered in 
the non-centrosymmetic ferromagnetic compound UIr under pressure \cite{UIr}. The symmetry 
of the SC order parameter and its connection to FM nevertheless remains unclear, which 
motivates the present study. Spontaneous symmetry breaking in condensed matter systems 
is conceptually of immense importance, as it may provide clues for what could be expected 
in systems belonging to vastly different areas of physics. The study of a condensed-matter 
system such as UIr with multiple broken symmetries is likely to have impact on a number of 
disciplines of physics, including such disparate phenomena as mass differences between 
elementary particles and extremely dilute ultra-cold atomic gases. 
\par
In this work, we study a model system of a non-centrosymmetric superconductor with 
substantial spin-orbit coupling, which at the same time exhibits itinerant ferromagnetism.
The origin of the SOC may be either that the crystal structure lacks a center of inversion, 
such as in UIr, or due to a thin-film geometry where the breakdown of inversion symmetry 
near the surface induces transverse electrical fields, leading to the well-known Rashba
SOC~\cite{rashba}. Our model should therefore be relevant both to the non-centrosymmetric and
centrosymmetric heavy fermion compounds, since the SOC is considerable in any case due to
the high atomic number. Specifically, materials that exhibit coexistence of SC and FM order 
and where SOC is large include UGe$_2$~\cite{saxena}, URhGe~\cite{URhGe}, UCoGe~\cite{UCoGe}, 
and UIr~\cite{UIr}. For this model, we construct a  mean-field theory, solve the saddle 
point equations for the order parameters and study the effect of spin-orbit coupling on the superconducting 
order parameters. Finally, we discuss  application of this model to the case of UIr.
\par
To label the SOC+FM split bands, it is possible to introduce a \emph{pseudospin} basis in which the normal-state Hamiltonian is diagonalized. In the original spin basis, the SC matrix
order parameter is characterized, in analogy to the $p$-wave state~\cite{leggett1975}, by a vector
$\mathbf{d}_\kk$ and scalar $\Delta_s$ so that $\hat{\Delta}_{\alpha\beta}(\kk) = \i\Delta_s\hat{\sigma}_y + 
[\i(\mathbf{d}_\kk\cdot\hat{\boldsymbol{\sigma}}) \hat{\sigma}_y ]_{\alpha\beta}$.
Note that, unlike the usual $p$-wave 
SC, a singlet component $\Delta_s$ of the gap will also be present since antisymmetric SOC in general mixes the 
parity of the order parameter. Below, $\hat{\ldots}$ is used for $2\times2$ matrices.
We now proceed to write down the effective Hamiltonian $H = H_\text{N} + H_\text{SC}$ for 
our system. In the normal state, the  Hamiltonian in momentum-space reads \cite{prbcoexistence}
\begin{align}
H_\text{N} = H_0 +  \sum_{\vk\alpha\beta}\, [c_{\vk\alpha}^\dag(\varepsilon_\vk\hat{1} - h\hat{\sigma}_z + \hat{\boldsymbol{\sigma}}\cdot\mathbf{g}_\vk )_{\alpha\beta}c_{\vk\beta}],
\end{align}
where $H_0=INM^2/2$.
Above, the dispersion relation $\varepsilon_\vk$ is measured from chemical potential $\mu$, and the 
magnetization $M = |\mathbf{M}|$ is taken along the easy-axis, while $h=IM$ is the 
exchange splitting of the bands and $\mathbf{g_k}$ is the SOC vector.
When superconductivity coexists with FM, the SC pairing is generally believed to be non-unitary 
\cite{samokhin}, characterized by $\mathbf{d}_\vk\times \mathbf{d}_\vk^* \neq 0$. In such a 
scenario, the SC order parameter couples to the spontaneous magnetization $\mathbf{M}$ through a 
term $\gamma\mathbf{M}\cdot\mathbf{d}_\vk\times\mathbf{d}_\vk^*$ in the free energy, where the sign 
of $\gamma$ is determined by the gradient of the DOS at Fermi level \cite{machida} and $\langle
\mathbf{S}_\vk \rangle = \i\mathbf{d}_\vk\times\mathbf{d}_\vk^*$ is the spin associated 
with the Cooper pair. Thus, for $\gamma<0$ it is expected that a SC pairing state obeying
$\i\mathbf{d}_\vk\times\mathbf{d}_\vk^* \parallel \mathbf{M}$ is energetically favored, implying that 
$\mathbf{d}_\vk$ must be complex-valued. Our model captures broken
time-reversal symmetry in addition to antisymmetric
SOC. As shown by Anderson~\cite{anderson}, the presence of the latter is detrimental to spin-triplet 
SC pairing state, unless $\mathbf{d}_\vk \parallel \mathbf{g}_\vk$. 
In our case, it is obvious that a
non-unitary SC pairing state cannot satisfy this condition  
since $\dd_\kk$ is complex, whereas $\mathbf{g}_\kk$ must be real for the Hamiltonian to be hermitian.

 The SOC vector reads $\mathbf{g}_\vk = -\mathbf{g}_{-\vk}$, and we 
introduce $g_{\vk} = \mathbf{g}_{\vk,x} - \i\mathbf{g}_{\vk,y}$ for
later use. We consider the SOC in the Rashba form, namely  $\mathbf{g}_\vk = \lambda(k_y,-k_x,0)$. This cor\-responds to a
situation where an asymmetric potential gradient is present along the $\mathbf{z}$-axis, and is 
also the scenario realized in non-centrosymmetric CePt$_3$Si~\cite{frigeri04}. We have introduced 
fermion operators $\{c_{\vk\sigma}\}$ in a basis 
$\varphi_\vk = [c_{\vk\uparrow}, c_{\vk\downarrow}]^\mathrm{T}$.  
\par
Diagonalizing the normal-state Hamiltonian yields the  quasiparticle excitations $\tilde{E}_{\vk\sigma}
= \varepsilon_\vk - \sigma\sqrt{h^2+\lambda^2k^2}$, which due to the SOC 
are characterized by the pseudospin $\sigma=\pm1$. For later use, we define $\mathcal{N}_\vk = [1+\lambda^2k^2/(h + 
\sqrt{h^2+\lambda^2k^2})^2]^{-1/2}$.
The superconducting pairing is now assumed to occur between the excitations described by
$\tilde{\varphi}_\vk$. Due to the presence of antisymmetric spin-orbit coupling, 
this automatically leads to a mixed-parity SC state in the original spin basis. To see this, we
introduce 
\begin{align}
H_\text{SC} = \frac{1}{2N} \sum_{\vk\vk'\sigma}
V_{\vk\vk'\sigma} \tilde{c}_{\vk\sigma}^\dag
\tilde{c}_{-\vk\sigma}^\dag \tilde{c}_{-\vk'\sigma}
\tilde{c}_{\vk'\sigma},
\end{align}
and perform a standard mean-field decoupling,
which after an additional diagonalization yields the total Hamiltonian
in the superconducting state: $H=H_0+\Sigma_{\vk\sigma} (\tilde{E}_{\vk\sigma}-E_{\vk\sigma}-\tilde{\Delta}_{\vk\sigma} \tilde{b}_{\vk\sigma}^\dag+2\eta_{\vk\sigma}^\dag \eta_{\vk\sigma})/2,$
 where $E_{\vk\sigma} =
(\tilde{E}_{\vk\sigma}^2 + |\tilde{\Delta}_{\vk\sigma}|^2)^{1/2}$ and
$\{\eta_{\vk\sigma}^\dag,\eta_{\vk\sigma}\}$ are new fermion
operators. The merit of this procedure is that we can now obtain simple
self-consistency equations for the gaps $\tilde{\Delta}_{\vk\sigma}$,
which may then be transformed back to the gaps in the original
spin-basis $\varphi_\vk$ by means of the unitary transformation
$P_\vk$. We assume a chiral
$p$-wave symmetry for the gaps with a corresponding pairing potential
$V_{\vk\vk'\sigma} = -g_{sc}\e{\i\sigma(\phi-\phi')}$, where $\tan\phi =
k_x/k_y$. The motivation for this is that this choice ensures that the 
condition $\mathbf{d}_\vk \parallel \mathbf{g}_\vk$ is satisfied exactly 
for $h\to 0$, and corresponds to a fully gapped Fermi surface which favors 
the condensation energy. The gaps obtain the form $\tilde{\Delta}_{\vk\sigma} =
-\sigma \tilde{\Delta}_{\sigma,0} \e{\i\sigma\phi}$ and we find a self-consistency 
equation of the standard BCS form with a cutoff $\omega$ on the pairing-fluctuation spectrum which we do not 
specify further. Moreover, $N^\sigma(\varepsilon)$ is the pseudospin-resolved 
density of states (DOS) for the $\tilde{E}_{\vk\sigma}$ ($\sigma$=$\pm$) bands of 
the quasiparticle excitations~\cite{foot2}.
Introducing the total DOS at the Fermi level for a normal metal $N_0 = mV\sqrt{2m\mu}/\pi^2$ 
and defining $c=gN_0/2$, the analytical solution for the gaps reads 
$\tilde{\Delta}_{\sigma,0} = 2\omega\text{exp}\{-1/[cR_\sigma(0)]\},\; R_\sigma(\varepsilon) = 2N^\sigma(\varepsilon)/N_0$.
With the analytical solution for $\tilde{\Delta}_{\sigma,0}$ in hand, we may exploit the unitary 
transformation $P_\vk$ to express the superconducting gaps in the original spin basis as follows:
\begin{align}\label{eq:gaps}
\Delta_{\vk\uparrow} &= -\e{\i\phi}[\tilde{\Delta}_{\uparrow,0} (\mathcal{N}_\vk^\uparrow)^2  + \tilde{\Delta}_{\downarrow,0}(\mathcal{N}^\downarrow_\vk)^2\lambda^2 k^2_\downarrow(0)\Lambda^2_{\vk\downarrow} ],\notag\\
\Delta_{\vk\downarrow} &= \e{-\i\phi}[\tilde{\Delta}_{\downarrow,0} (\mathcal{N}_\vk^\downarrow)^2  + \tilde{\Delta}_{\uparrow,0}(\mathcal{N}^\uparrow_\vk)^2\lambda^2 k^2_\uparrow(0)\Lambda^2_{\vk\uparrow} ],\notag\\
\Delta_{\vk\uparrow\downarrow} &= -\sum_\sigma \tilde{\Delta}_{\sigma,0}(\mathcal{N}^\sigma_\vk)^2\lambda |k_\sigma(0)|\Lambda_{\vk\sigma},\; \sigma=\pm 1,
\end{align}
where we have defined $\mathcal{N}^\sigma_\vk = \mathcal{N}_{\vk=\vk^\sigma(0)}$ and $\Lambda_{\vk\sigma} = [h + \sqrt{h^2 + \lambda^2k^2_\sigma(0)}]^{-1}$. Note that in the original spin 
basis, the superconducting order parameter is in general a mixture of triplet $(\Delta_{\vk\sigma})$ and singlet 
$(\Delta_{\vk\uparrow\downarrow})$ components. The self-consistency equation for the magnetization is:
\begin{align}\label{eq:mag}
h + \frac{\tilde{I}}{4} \sum_\sigma \int  \frac{\sigma \text{d}\varepsilon R^\sigma(\varepsilon) h\varepsilon}{\sqrt{[h^2+\lambda^2k_\sigma^2(\varepsilon)](\varepsilon^2 + \tilde{\Delta}_{\sigma,0}^2)}} = 0,
\end{align}
where the integration is over the bandwidth and
$\tilde{I}=IN_0$. Eqs. (\ref{eq:gaps}, \ref{eq:mag}) 
are the main analytical results of this work.
\par
Let us briefly investigate some important limiting cases of Eq. (\ref{eq:gaps}). In the absence of 
spin-orbit coupling $(\lambda\to0)$, one finds $\mathcal{N}^\sigma_\vk \to 1$ and $\Delta_{\vk\sigma} 
= \tilde{\Delta}_{\vk\sigma}$ while $\Delta_{\vk\uparrow\downarrow} = 0$, such that we reproduce the 
results of Refs. \cite{nevidomskyy,prbcoexistence}. In the absence of an exchange energy $(h\to0)$, 
one finds that $\mathcal{N}^\sigma_\vk \to 1/\sqrt{2}$ and 
$\Delta_{\vk\uparrow} = 
-\e{\i\phi}(\tilde{\Delta}_{\uparrow,0} + \tilde{\Delta}_{\downarrow,0})/2$, $\Delta_{\vk\downarrow} 
= \e{-\i\phi}(\tilde{\Delta}_{\uparrow,0} + \tilde{\Delta}_{\downarrow,0})/2$, and 
$\Delta_{\vk\uparrow\downarrow} = (\tilde{\Delta}_{\downarrow,0} - \tilde{\Delta}_{\uparrow,0})/2$. 
As demanded by consistency, the triplet gaps are equal in magnitude since there is no exchange field 
and the singlet component is nonzero since $\tilde{\Delta}_{\uparrow,0} \neq \tilde{\Delta}_{\downarrow,0}$ 
in general. Finally, Eq. (\ref{eq:mag}) reproduces the well-known Stoner criterion 
$\tilde{I} \geq1$ for the onset of FM in the absence of SOC and SC $(\lambda\to 0, g\to0)$. 
\par
\begin{figure}[t]
\centering
\resizebox{0.48\textwidth}{!}{
\includegraphics{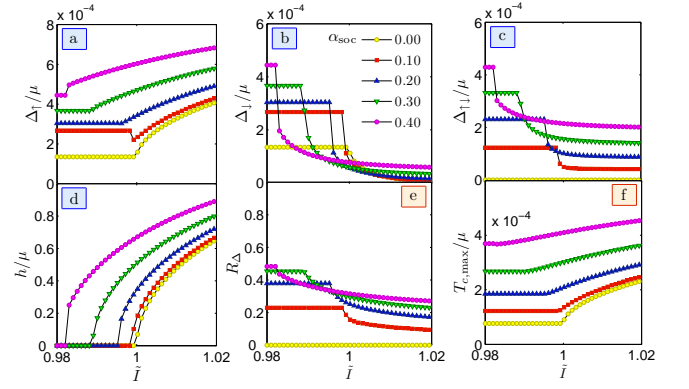}}
\caption{(Color online) Self-consistent solution of the order parameters (a-d) as a function of the FM exchange 
	parameter $\tilde{I}$, the ratio between the singlet and triplet gaps (e) $R_\Delta = \Delta_{\uparrow\downarrow}/(\Delta_\uparrow+\Delta_\downarrow)$ and the maximal critical temperature (f)
$T_{c,\mathrm{max}}$ as a function of the FM exchange parameter $\tilde{I}$.}
\label{fig:one} 
\end{figure}
\par
We now focus on the general case in which $h\neq0$ and $\lambda\neq0$. First of all, we must specify the range of the parameters 
in the problem that corresponds to a physically realistic scenario. We allow $h$ to range, in principle, from 0 to $\mu$, the 
latter denoting a fully polarized ferromagnet. 
As a convenient measure of the 
strength of SOC, we introduce the dimensionless quantity $\alpha_\text{soc} \equiv \sqrt{2\lambda^2 m/\mu}$ which has a 
direct physical interpretation: namely, it is the ratio of the SOC (at
$E_F$) to the Fermi energy $\mu$.
The parameter  $\alpha_\text{soc}$  is allowed to vary from 0 to 
$\delta$, where $\delta$ denotes a fraction of the Fermi energy. We take $\delta=0.5$ as a sensible upper limit. 
Note that generically, 
the SOC strength  at the Fermi level is different for the two quasiparticle bands, and moreover depends on $h$. For a 
given value of $h$, one may derive that $\lambda\leq\delta\mu/[2\mu m + \sqrt{2m^2(h^2+\delta^2\mu^2)}]^{1/2}$ ensures that 
the spin-orbit energy is less than $\delta\times\mu$ for both quasiparticle bands. 
\par
In Fig. \ref{fig:one}a-d, we present the self-consistent solutions for the order parameters in Eqs. (\ref{eq:gaps}) 
and (\ref{eq:mag}) as a function of the FM exchange parameter $\tilde{I}$ for several values of $\alpha_\text{soc}$. 
We have defined $\Delta_{\sigma}=|\Delta_{\vk\sigma}|$ and $\Delta_{\uparrow\downarrow} = |\Delta_{\vk\uparrow\downarrow}|$, 
and fixed $\omega/\mu = 0.01$ and $m/\mu=5\times10^4$ with $c=0.2$, which are standard choices. For $\alpha_\text{soc}=0$, 
the onset of FM occurs at $\tilde{I}=1.0$ which lifts the degeneracy of $\Delta_\uparrow$ and $\Delta_\downarrow$, 
while $\Delta_{\uparrow\downarrow}$ is always zero. Upon increasing $\alpha_\text{soc}$, it is interesting to note 
that the PM-FM transition occurs at lower values of $\tilde{I}$, indicating that spin-orbit coupling favors  
ferromagnetic ordering. For $\alpha_\text{soc}\neq 0$, it is seen that $\Delta_{\uparrow\downarrow}$ is also non-zero, 
although it becomes suppressed at the onset of ferromagnetism. A common feature for all gaps is that they increase 
with $\alpha_\text{soc}$ in the absence of ferromagnetism \textit{and} deep inside the ferromagnetic phase 
$\tilde{I} \geq 1.02$. In the intermediate regime, there are crossovers between the gaps for different values 
of $\alpha_\text{soc}$ due to the different onsets of ferromagnetic order. By comparing the behaviour between 
the gaps for increasing $\tilde{I}$ with $\alpha_\text{soc}\neq 0$, one infers that $\Delta_\downarrow$ and 
$\Delta_{\uparrow\downarrow}$ eventually saturate at a constant non-zero value, while $\Delta_\uparrow$ continues 
to increase steadily. This is quite different from the case when $\alpha_\text{soc}=0$, where the minority 
spin-gap goes to zero rapidly with increasing $\tilde{I}$. This seems to suggest that the presence of spin-orbit 
coupling in the system ensures the survival of the minority-spin gap $\Delta_\downarrow$ and the singlet gap 
$\Delta_{\uparrow\downarrow}$ even though the FM exchange energy becomes strong. 
\par
In Fig. \ref{fig:one}e and \ref{fig:one}f, we plot the ratio of the singlet and triplet gaps, defined as 
$R_\Delta = \Delta_{\uparrow\downarrow}/(\Delta_\uparrow+\Delta_\downarrow)$, and the maximal 
critical temperature $T_{c,\mathrm{max}}$ 
for the onset of superconductivity. It is seen from the left panel that $R_\Delta$ increases with 
$\alpha_\text{soc}$ in the PM regime, suggesting that the singlet component becomes more prominent 
in the system as compared to the triplet gaps. However, at the onset of FM order, 
$R_\Delta$ decreases since the singlet component becomes suppressed by the Zeeman-splitting. 
In the right panel, one observes that $T_{c,\text{max}}$ increases both with $\alpha_\text{soc}$ 
and $\tilde{I}$. Our findings suggest that the presence of
antisymmetric SOC, originating from 
\eg non-centrosymmetricity of the crystal structure,  \textit{enhances} \textit{both} the tendency 
towards ferromagnetism and the magnitude of the SC gaps in all spin channels. In the absence 
of spin-orbit coupling, it was shown in Ref. \cite{prbcoexistence} that the simultaneous 
coexistence of FM and non-unitary triplet superconductivity is the thermodynamically 
favored state as compared to the pure normal, FM, or SC state. Since the presence of spin-orbit 
coupling is seen to enhance both the FM and SC order parameters, it is reasonable to expect 
that the coexistent state is still thermodynamically the most favorable one even 
when $\alpha_\text{soc}\neq 0$. 
\par
Out of the known non-centrosym\-metric superconductors, UIr is the only
compound that is also a ferromagnet. This material, which is ferromagnetic at
ambient pressure, develops superconductivity in a narrow pressure
region around $P\sim2.6$~GPa right next to the FM--PM quantum phase transition, 
with a maximum SC transition temperature $T_\text{SC}\sim0.14$~K~\cite{UIr}. 
At this pressure, the saturated magnetic moment was measured to be 
0.07$\mu_B$ per U atom, and such a small value clearly indicates the 
itinerant character of the ferromagnetism, presumably due to 5$f$ 
electrons of uranium. UIr crystallizes in the monoclinic structure 
(space group $P$2$_1$) which lacks inversion symmetry, and the FM 
moment is Ising-like, oriented along the [10$\bar{1}$] direction 
in the (ac)-plane. 
\par
Given the proximity of the SC state in UIr to the PM transition, one
may probably consider the magnetization $h$ as a perturbation on top of the
SOC-split bands. Neglecting the effect of the former, it is known~\cite{samokhin_CePt3Si} that 
even in the case of non-centrosymmetric superconductors (and $h=0$), the band energies
still satisfy the relation $\eps_\beta(\kk) = \eps_\beta(-\kk)$ due to the
time reversal symmetry of the single-electron Hamiltonian. As a
consequence, the SC order parameter on the $\beta$-th sheet of the
Fermi surface transforms according to one of the irreducible
representations of the normal state point group. In the case
of UIr, the point group $\mathbf{C_2}$ has two one-dimensional
irreducible representations, denoted A and B. Then the
SC order parameter is an odd function $\Delta(-\kk)=-\Delta(\kk)$ given by~\cite{samokhin_erratum_sergienko}
$\Delta^{A,B}_\beta(\kk)\, \propto\, t(\kk)\, \phi^{A,B}_\beta(\kk),$
where $t(\kk)$ is an odd phase factor~\cite{phase_factor} and the
basis functions $\phi^{A,B}$ are even in $\kk$.
Denoting the rotation axis of the $\mathbf{C_2}$ group as $z$
(this actually corresponds to $b$-axis in case of UIr), 
 the even functions $\phi_A$ and $\phi_B$ can then be cast in the following
form: $ \phi_A(\kk) = (k_z^2 + C)\, u_1(\kk)$, and
$\phi_B(\kk) = k_z (k_x\, u_2(\kk) + k_y\, u_3(\kk)) $,
where $C$ is some constant and $\{u_i(\kk)\}$ are arbitrary even functions of $k_x, k_y, k_z$.
Function $\phi_A$ generically has no nodes, whereas $\phi_B$ 
has two point nodes at the poles  \mbox{($k_x$=$k_y$=0)} and a line of nodes at the equator. 
The symmetry argument does not allow one to determine which pairing
channel is realized, however the experimental
observation of the strong pair-breaking effect due to disorder~\cite{UIr}
indicates that the gap must be anisotropic, possibly favouring the
gap with the nodes such as $\Delta_B(\kk)$.   
\par
One way of experimentally probing the symmetry of the superconducting order parameter in UIr would be by means of transport properties such as Josephson tunneling or point-contact spectroscopy. In particular, it has recently been shown that the presence of multiple gaps in superconductors with broken inversion symmetry should manifest itself through clear signatures at bias voltages corresponding to the sum and difference of the singlet and triplet components \cite{prbcoexistence,borkje,iniotakis}. We expect similar behavior in the present case, at least when the ferromagnetism is weak, and point-contact spectroscopy data could then be compared with the predictions for $R_\Delta$ in Fig. \ref{fig:one}e. Alternatively, it should be possible to directly probe the spin-texture of the superconducting 
order parameter by studying the effect of an externally applied magnetic field when 
the paramagnetic limitation dominates, \eg in a thin-film structure, 
where the orbital mechanism of destroying superconductivity is suppressed~\cite{Pauli-field}.
\par 
In summary, we have developed a mean-field model for a superconductor lacking inversion symmetry and displaying itinerant ferromagnetism. Specifically, we have investigated the interplay between ferromagnetism 
and asymmetric spin-orbit coupling and how these affect superconducting order, which in general is a 
mixture of a singlet and triplet components. Our main results are the analytical 
expression Eqs. (\ref{eq:gaps}) and (\ref{eq:mag}) and the belonging
discussion. 
We find that spin-orbit coupling may \textit{enhance} superconductivity 
in both the singlet and triplet channels in addition to favoring the Stoner criterion 
for the ferromagnetic instability. We have applied these considerations to the 
heavy fermion superconductor UIr, together with group-theo\-re\-tical
analysis of the symmetry of the SC order parameter.
\par
\textit{Acknowledgments.} The authors are grateful to M. Dzero for bringing
Ref.~\onlinecite{gorkov-rashba} to our attention. A.H.N. was supported by the US NSF grant DMR 0605935.
J.L. and A.S. were supported by the Norwegian Research Council 
Grants No. 158518/431 and No. 158547/431 (NANOMAT), and Grant No. 167498/V30 (STORFORSK).


\begin{thebibliography}{99}
\bibitem{phase_factor} The nontrivial phase factor $t(\kk)=-t(\kk)$ is
  defined such that $T\state{\kk} = t(\kk)\state{-\kk}$, where $T$ is
  the time-reversal operator.
\bibitem{Pauli-field} Experimentally, the measurements  
on a single crystal of UIr so far indicate that the orbital effects of the magnetic field dominate.
\end{thebibliography}
\end{document}